\documentclass[aps,prb,twocolumn,showpacs]{revtex4}
\usepackage{graphicx,bm,color,float,textcomp,mathtools}			
\newcommand\PRB[3]{Phys. Rev. B {\bf {#1}}, {#2} ({#3})}	
\newcommand\JPCM[3]{J. Phys.: Cond. Matter {\bf {#1}}, {#2} ({#3})}
\newcommand\JPSJ[3]{J. Phys. Soc. Japan {\bf {#1}}, {#2} ({#3})}

\newcommand\PRL[3]{Phys. Rev. Lett. {\bf {#1}}, {#2} ({#3})}					
\newcommand\JPCS[3]{J. Phys.: Conf. Series {\bf {#1}}, {#2} ({#3})}

\newcommand\seo{$\rm SrEr_2O_4$}
\newcommand\sdo{$\rm SrDy_2O_4$}
\newcommand\sho{$\rm SrHo_2O_4$}
\newcommand\syo{$\rm SrYb_2O_4$}
\newcommand\sgo{$\rm SrGd_2O_4$}
\newcommand\slo{$\rm Sr {\it Ln}_2O_4$}

\begin{document}

\title{Low-Temperature Magnetism in the Honeycomb Systems \slo}
\author{O.A. Petrenko}
\affiliation{University of Warwick, Department of Physics, Coventry, CV4~7AL, UK}
\date{\today}

\begin{abstract}
Recent progress in the understanding of the complex magnetic properties of the family of rare-earth strontium oxides, \slo, is reviewed.
These compounds consisting of hexagons and triangles are affected by geometrical frustration and therefore exhibit its characteristic features, such as a significant reduction of magnetic ordering temperatures and complex phase diagrams in an applied field.
Some of the observed features appear to be rather remarkable even in the context of the unusual behaviour associated  in geometrically frustrated magnetic systems.
Of particular interest is the coexistence at the lowest temperature of different magnetic structures (exhibiting either long or short-range order) characterised by different propagation vectors in materials without significant chemical or structural disorder.
\end{abstract}

\maketitle
\section{Introduction}
Frustrated magnets have been a focal point of the research on magnetism for the past two decades.
In this article, the influence of geometrical frustration on the magnetic properties of the family of rare-earth strontium oxides, \slo, (where $Ln=$Gd, Dy, Ho, Er, Tm, and Yb) is discussed. 
Given the nature of this special issue of {\it Low Temperature Physics} on antiferromagnetism an extensive general introduction to magnetically frustrated systems is omitted and the reader is instead referred to a collection of reviews available on the subject~\cite{Frustration}.
We start with a description of the crystal structure and general properties of \slo\ and other closely related compounds and then present the recently obtained experimental results on their low-temperature magnetic properties by our group and others.
Particular attention is paid to the zero-field ground state of \seo, \sho, and \sdo\ (section~\ref{section_zero_field}), as well as the field-induced behaviour of these compounds (section~\ref{section_field}).
The penultimate section briefly reviews the other \slo\ compounds and discusses the importance of crystal field effects. 
The concluding section compares different members of the family and includes a brief summary.
\begin{figure}[tb]
\begin{center}
\includegraphics[width=0.8\columnwidth]{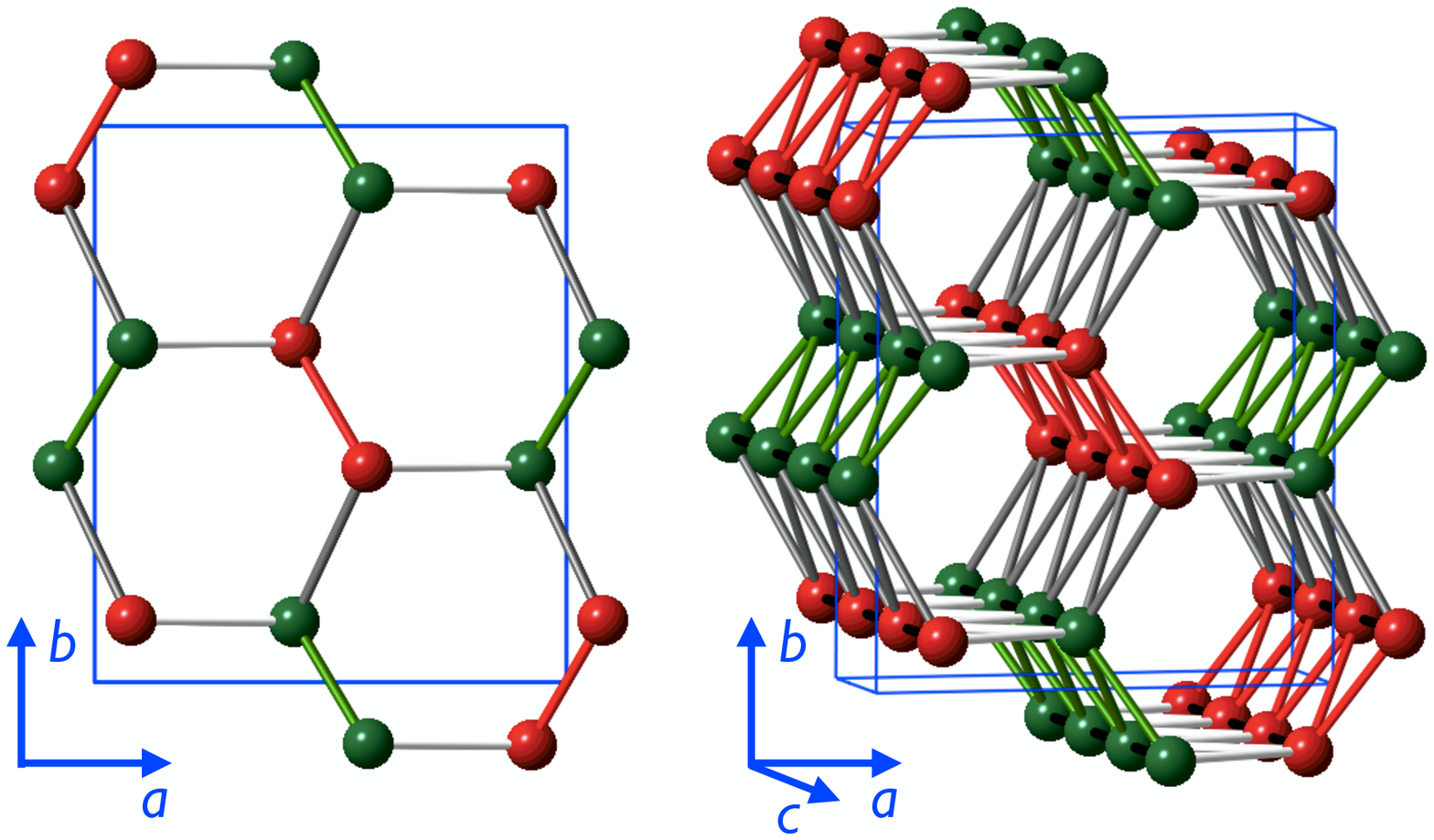}
\caption{\label{Fig1_Structure}
Positions of the magnetic rare-earth $Ln^{3+}$ ions within the \slo\ compounds, with the two crystallographically inequivalent $Ln^{3+}$ sites shown in different colours.
The left-hand panel emphasises the honeycomb arrangement of the $Ln^{3+}$ ions when viewed along the $c$ axis, while the right-hand panel demonstrates the formation of zigzag ladders running along the $c$ axis which link the honeycomb layers and give rise to geometric frustration.
The blue box represents a crystal unit cell of the $Pnam$ space group.}
\end{center}
\end{figure}

The members of the \slo\ family crystallise in the form of calcium ferrite~\cite{CalciumFerrite}, with the space group $Pnam$; the crystal structure of these materials (see Fig.~\ref{Fig1_Structure}) can be viewed as a network of linked hexagons and triangles~\cite{Lopato,Karun}.
The most important feature of the linked hexagon (or ``{\it honeycomb}") lattice is that it has the lowest coordination number, 3, in two dimensions.
This feature attracts a lot of theoretical attention to the lattice, but being bipartite, the honeycomb lattice is not frustrated if only the nearest-neighbour interactions are considered.
The frustration in a honeycomb lattice can be induced by further neighbour interaction and numerous models of frustrated honeycomb lattices which include Heisenberg, $XY$ or Ising $J_1-J_2-J_3$ interactions have been extensively studied theoretically, particularly for the $s=1/2$ quantum case.

In the \slo\ family, however, the cause of frustration is different; it arises from the triangular (or ``zigzag") ladders running along the $c$ axis which link the honeycomb layers.
In this respect \slo\ compounds are similar to recently reported $\beta-\rm CaCr_2O_4$~\cite{Damay} and perhaps to $Ln \rm V_4O_8$ compounds~\cite{Das}.
The term ``zigzag" has also been used to describe the spin-chain structure of another honeycomb lattice compound $\rm Na_2IrO_3$~\cite{Ye,Chaloupka}, but in a different context - to describe the arrangement of magnetic moments formed there.

The orthorhombic unit cell of the \slo\ compounds contains 4 Sr atoms on a single site, 8 $Ln$ atoms (split equally between two sites) and 16 oxygen atoms occupying 4 sites; all sites are of the $4c$ type with the coordinates $(x,y,1/4)$~\cite{Karun}.
The $a$ and $b$ axes of the unit cell are typically quite long, about 10 and 12~\AA\ respectively, while the $c$ axis is the shortest, at around 3.4~\AA\ on average.
The magnetic $Ln$ atoms are surrounded by the distorted oxygen octahedra, forming the chains running along the $c$-axis.
The shortest  $Ln-Ln$ separation is along the chains; there is a slightly larger separation between the chains formed by the $Ln$ atoms occupying the same sites (which are shown in Fig.~\ref{Fig1_Structure} in the same colour), while the distance between $Ln$ atoms from different sites (red and green in Fig.~\ref{Fig1_Structure}) is much greater~\cite{Karun}.
Such a crystal structure predetermines the quasi one-dimensional magnetic properties of the \slo\ compounds, as for rare-earth ions in insulating materials direct exchange is often the most important mechanism for magnetic coupling.
It is rather useful to note the equivalence of the well-studied linear chain model with nearest and next-nearest interactions~\cite{Yoshimori} and the ladders of rare-earth ions described here if these were ``stretched'' along the $c$ axis.

\begin{figure}[tb]
\begin{center}
\includegraphics[width=0.4\columnwidth]{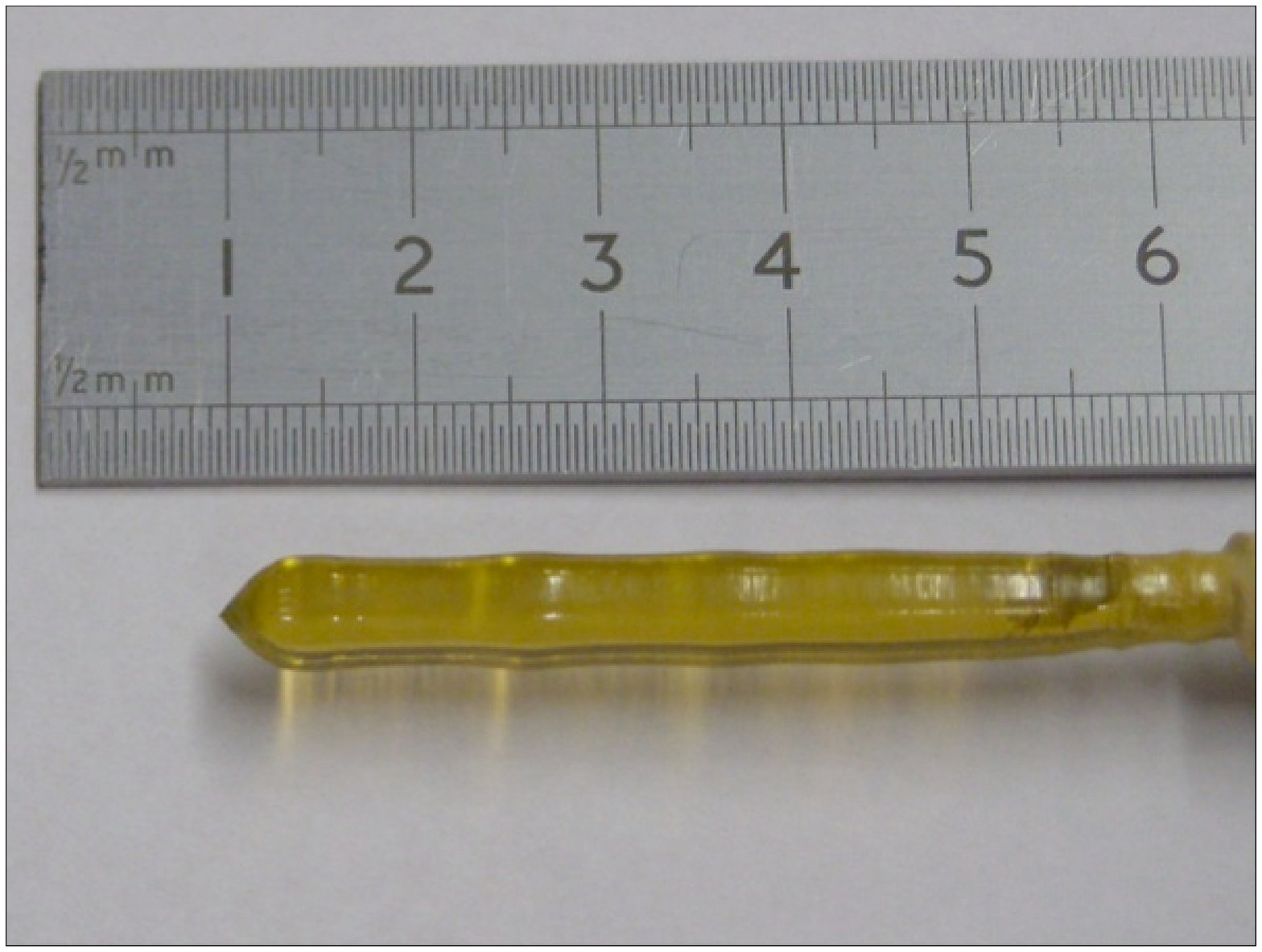}
\includegraphics[width=0.4\columnwidth]{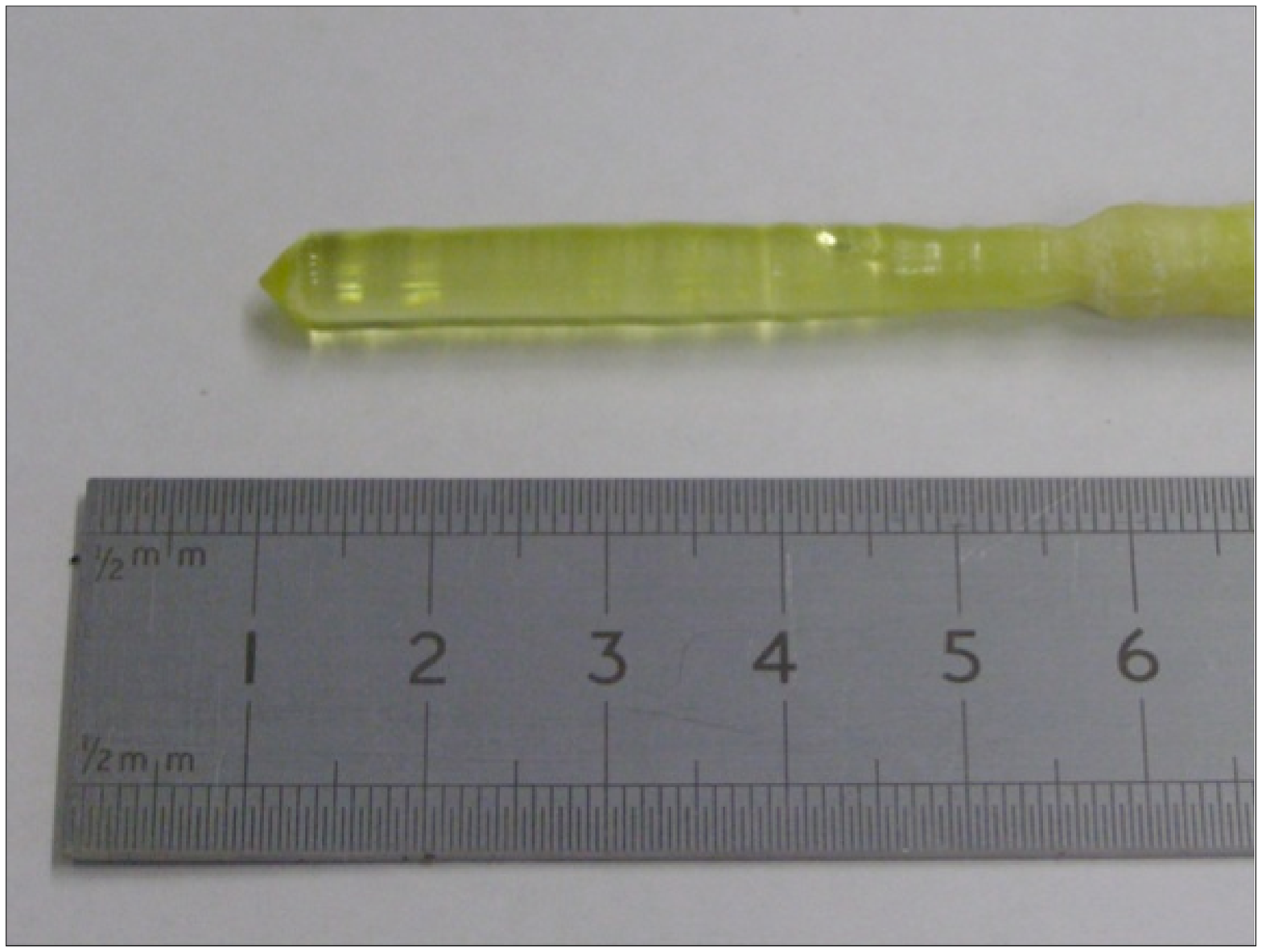}
\caption{\label{Fig2_Xtals}
As grown boules of \sho\ (left panel) and \sdo\ (right panel) single crystals, using growth speeds of 6 to 8~mm/h.
Figure is from Ref.~\cite{GB_JPCM_2009}.}
\end{center}
\end{figure}

An important observation to make prior to the description of their properties is that both polycrystalline and single crystal samples of the \slo\ compounds have been used for investigations. 
The progress achieved to date in the understanding of their complex behaviour is, however, largely due to the availability of high quality single crystals.
Crystals of magnetic \slo\ oxides and their non-magnetic analogues (with $Ln=$ Lu or Y) have been synthesised by the floating zone technique by our group and others~\cite{GB_JPCM_2009,Castro,Ghosh}.
Examples of the single crystals grown~\cite{GB_JPCM_2009} are shown in Fig.~\ref{Fig2_Xtals}; the size of the crystals available is certainly sufficient for neutron scattering experiments, including inelastic studies.
In comparison, the magnetic properties of structurally similar $\rm Ba{\it Ln}_2O_4$ (Ref.~\onlinecite{Doi}), $\rm Eu{\it Ln}_2O_4$ (Ref.~\onlinecite{Hirose,Ofer}) and $\rm Ba{\ it Ln}_2S_4$ (Ref.~\onlinecite{Misawa}) compounds have not yet been probed to a significant degree, as only polycrystalline samples are available.

\section{Zero field magnetic properties of \seo, \sho\ and \sdo}
\label{section_zero_field}
\subsection{\seo}
\begin{figure}[tb]
\begin{center}
\includegraphics[width=0.8\columnwidth]{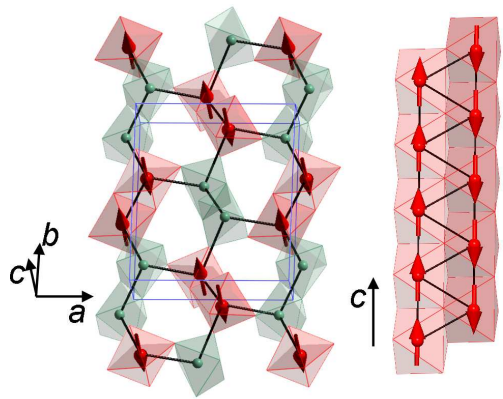}
\caption{\label{Fig3_SEO_str}
Magnetic structure of \seo\ as determined from Rietveld refinements of the neutron-diffraction pattern at $T=0.55$~K.
The same structure is shown twice to emphasise different arrangements of the magnetic moments along the $c$ axis and with respect to the hexagons in the plane.
Two different Er sites and their surroundings are shown in different colours.
Only one of the sites carries a significant magnetic moment.
Figure is from Ref.~[\onlinecite{OAP_PRB_2008}].}
\end{center}
\end{figure}

\begin{figure*}[tb]
\begin{center}
\includegraphics[width=1.5\columnwidth]{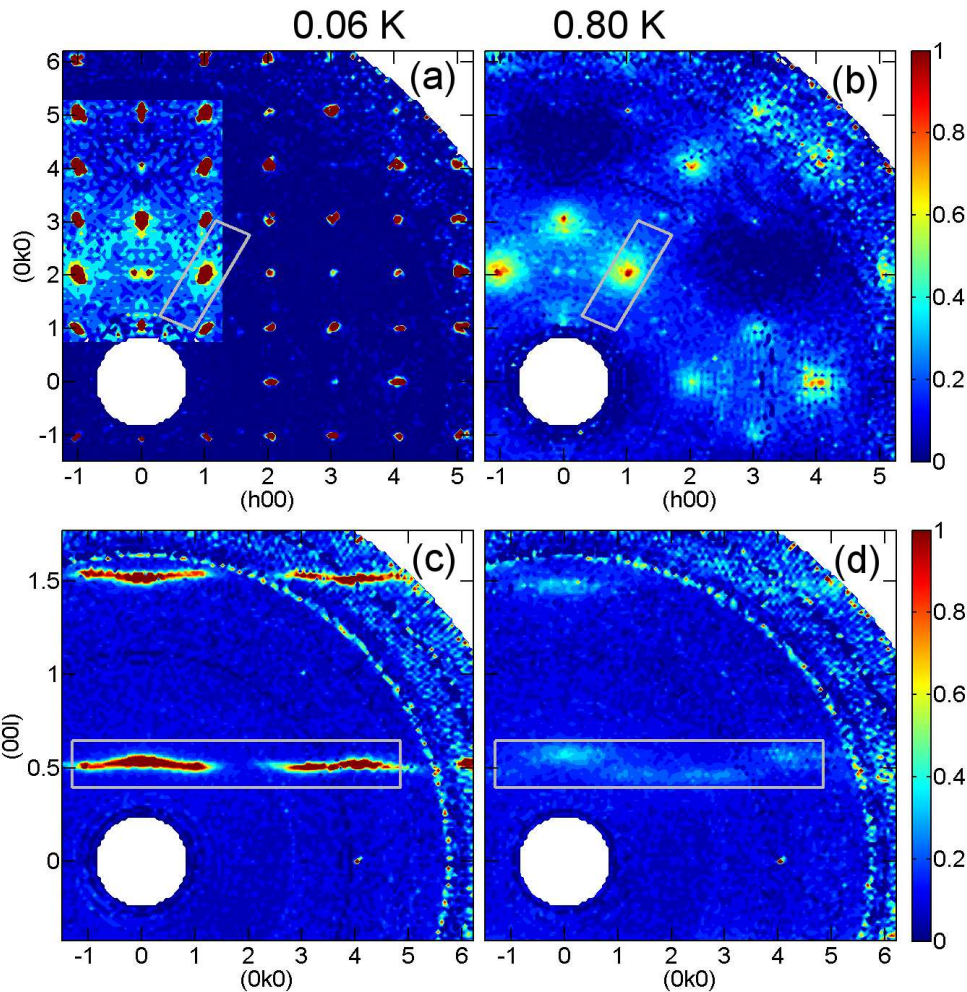}
\caption{\label{Fig4_SEO_diff}
Reciprocal space intensity maps of the magnetic scattering from \seo\ in the $(hk0)$ plane (top panels) and in the $(0kl)$ plane (bottom panels) at 0.06 K~(left panels) and 0.8~K (right panels).
The highlighted area $1.3 < h < 1.3$, $0.7 < k < 5.3$ in panel (a) has 10 times lower intensity scale to emphasise the presence at the lowest temperature of a weak diffuse scattering otherwise obscured by the much more intense Bragg peaks.
The magnetic scattering is isolated from the nuclear and spin-incoherent contribution by full XYZ polarisation analysis using D7 diffractometer for the $(hk0)$ plane.
In the case of maps of the $(0kl)$ plane, the intensity shown is obtained by removing the nuclear contribution from the non-spin-flip measurement with neutrons polarised orthogonal to the scattering plane, following Ref.~\cite{Stewart}.
Figure is from Ref.~[\onlinecite{TJH_PRB_2011}].}
\end{center}
\end{figure*}

\seo\ is found to order magnetically at $T_N=0.75$~K with a ${\bf k}=0$ antiferromagnetic (AFM) structure (depicted in Fig.~\ref{Fig3_SEO_str}) consisting of ferromagnetic chains running along the $c$ axis, with adjacent chains arranged antiferromagnetically~\cite{OAP_PRB_2008}.
The refinement of the powder neutron diffraction (PND) data suggested that the moments point along the $c$ direction and that only one of the two Er$^{3+}$ sites possesses a sizeable magnetic moment.
It was not possible to determine which particular site contributed to the ordering, as  the magnetic moments may be swapped between the two sites without changing the calculated PND pattern significantly~\cite{OAP_PRB_2008}.

The situation with the low-temperature magnetic structure of \seo\ became much clearer after the publication of single crystal polarised neutron diffraction results~\cite{TJH_PRB_2011}, which are summarised in Fig.~\ref{Fig4_SEO_diff}.
The presence of a magnetic component with long-range order (LRO) below 0.75~K was confirmed by the observation of sharp resolution-limited Bragg peaks at integer $(hkl)$ positions (see Fig.~\ref{Fig4_SEO_diff}a).
These peaks are replaced by broad and much weaker diffuse scattering features above $T_N$ (see Fig.~\ref{Fig4_SEO_diff}b).
Surprisingly, another distinct magnetic component corresponding to a short-range incommensurate structure was also detected.
This component manifests itself by the presence of a strong diffuse signal, forming the undulated planes of scattering, which are seen as ``{\it rods}" in a particular scattering plane.
Fig.~\ref{Fig4_SEO_diff}c, for example, clearly shows the two rods are at positions $(0,k,1/2+\delta)$ and $(0,k,3/2-\delta)$, where $\delta$ is dependent upon $k$.
A Monte Carlo simulation~\cite{TJH_PRB_2011} showed that a simple model based on a ladder of triangles in which the nearest-neighbour interactions are approximately five times stronger than the next-nearest-neighbour interactions satisfactorily mimics the observed diffuse scattering patterns.

From the width of the diffuse ``{\it rods}" at the base temperature, the estimates for correlation length along the $c$ axis vary from 130 to 70~\AA\ depending on which ``{\it rod}" is considered~\cite{TJH_PRB_2011}, but in any event the AFM correlations are rather long and include more than 20 magnetic ions.
The interpretation of these data is that apart from the ${\bf k}=0$ LRO component shown in Fig.~\ref{Fig3_SEO_str}) the magnetic structure of \seo\ consists of highly correlated AFM chains running along the $c$ axis, but the correlations between the chains are rather weak.

On warming from the base temperature of a dilution cryostat to much higher temperatures, the partially ordered component does not undergo a pronounced phase transition unlike the ${\bf k}=0$ component.
Instead it gradually loses the intensity, but it could be easily seen at 0.8~K (see Fig.~\ref{Fig4_SEO_diff}d) and in fact much higher temperatures (not shown).

From the polarisation analysis~\cite{TJH_PRB_2011}, the magnetic moments in the long-range commensurate and short-range incommensurate structures are found to be predominantly pointing along the $c$ and $a$ axes, respectively.

\subsection{\sho}
At a first glance, a refinement of the low-temperature magnetic structure of \sho\ looks very similar to \seo.
The PND data~\cite{OY_JPCS_2012} returned a collinear AFM ${\bf k}=0$ component, very similar to the one shown in Fig.~\ref{Fig3_SEO_str}, below the ordering temperature of 0.68~K, with only a half of the Ho$^{3+}$ ions carrying a significant moment.
The presence of another magnetic component was also observed as a pronounced scattering around the $(0,0,1/2)$ positions.
Further single crystal diffraction data~\cite{OY_PRB_2013}, however, revealed a more complicated picture.

The observed broad diffraction peaks show that the ${\bf k}=0$ component (corresponding to a collinear antiferromagnetically coupled structure) is of short-range order type.
The planes of diffuse scattering corresponding to another kind of magnetic order appear to be nearly perfectly commensurate, {\it i.e.} the parameter $\delta$ is almost zero for them, although the variations of the intensity have been seen in both the $(h0l)$ and $(0kl)$ planes in reciprocal space.
This observation suggests that the second type of short-range order present in \sho\ is principally one-dimensional in nature, that is the magnetic structure is essentially a collection of AFM coupled chains running along the $c$ axis with the intrachain correlations remaining rather weak down to lowest temperatures.
Similarly to what have been observed in \seo, a magnetic component with the propagation vector ${\bf k}=0$ exists below a well-defined transition temperature, while the one-dimensional scattering is observed at much higher temperatures.

Correlation lengths associated with the broad peaks are about 150~\AA\ in the $ab$ plane and about 190~\AA\ along the $c$ axis, while the correlation length associated with the diffuse scattering planes is 230~\AA\ along the $c$ axis at the lowest temperature.
From the polarisation analysis~\cite{OY_PRB_2013}, the magnetic moments in the ${\bf k}=0$ and quasi one-dimensional structures are found to be pointing along the $c$ and $b$ axes, respectively.

\subsection{\sdo}
\begin{figure}[tb]
\begin{center}
\includegraphics[width=0.8\columnwidth]{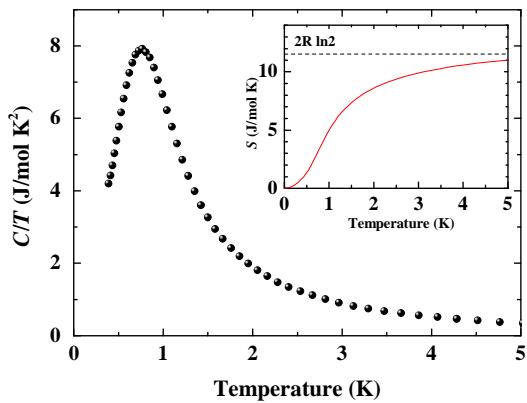}
\caption{\label{Fig5_SDO_CvsT}
Temperature dependence of the specific heat divided by temperature of \sdo\ in zero field.
The inset shows the temperature dependence of the entropy, $S$ (solid line), calculated as the area under the $C(T)/T$ curve which has been extended linearly down to $T=0$~K.
The dashed line indicates the position of 2R$\ln(2)$, which corresponds to the magnetic contribution for a system with an effective $s=1/2$.
Figure is from Ref.~[\onlinecite{THC_JPCM_2013}].}
\end{center}
\end{figure}

In contrast to the other members of the \slo\ family investigated so far, \sdo\ does not show any sign of magnetic phase transition down to the lowest available temperatures.
In zero field, heat capacity $C(T)$ measurements indicate that this compound appears to be magnetically disordered down to at least~0.39~K (see Fig.~\ref{Fig5_SDO_CvsT}).
The $C(T)/T$ curve shows a very broad maximum at 0.77~K and a nearly linear temperature dependence below this peak.
There are no sharp features in the heat capacity curve which can be attributed to a phase transition to a magnetically ordered state.
PND data for \sdo\ show no signs of any long-range magnetic order down to 20~mK, as the scattering pattern in zero field is dominated by broad diffuse scattering peaks~\cite{Petrenko_unpublished}.

The magnetic entropy recovered in \sdo\ between zero temperature and $T=5$~K (see inset in Fig.~\ref{Fig5_SDO_CvsT}) amounts to $2R\ln{2}$, which suggests that at the lowest temperature the system is essentially a doublet with the magnetic moments restricted to point only along the easy axis (Ising) direction.

\section{Field-induced behaviour of \seo, \sho\ and \sdo}
\label{section_field}
\begin{figure*}[tb]
\begin{center}
\includegraphics[width=1.5\columnwidth]{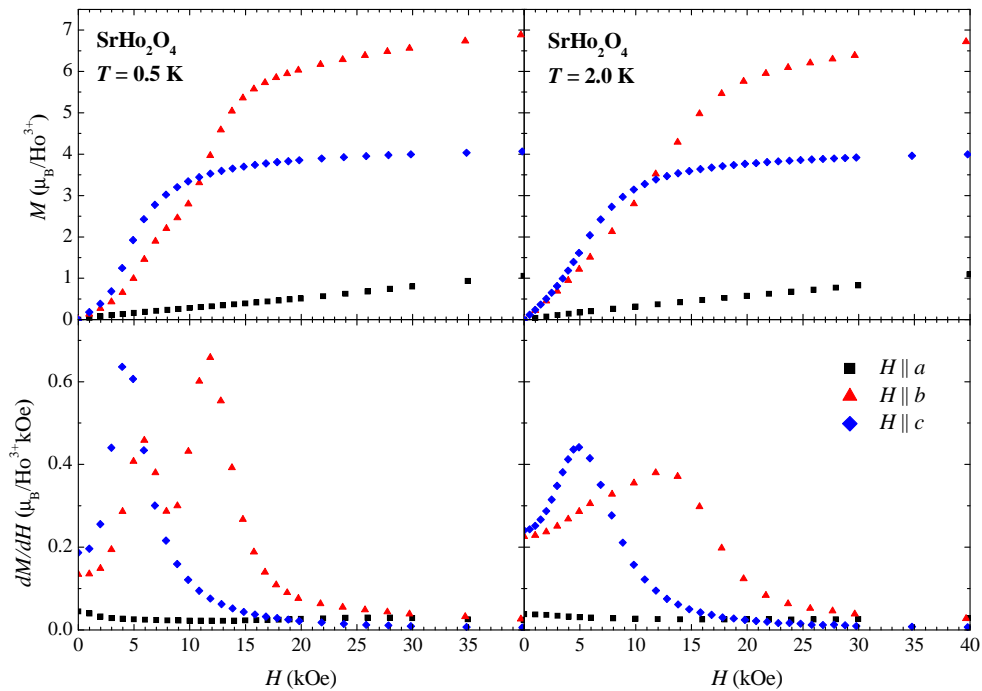}
\caption{\label{Fig6_SHO_MvsH}
Field-dependent magnetisation curves (top panels) for \sho\ obtained at 0.5~K (left) and 2.0~K (right) in the range of fields 0 to 40~kOe.
(Bottom panels) The field derivatives of the magnetisation at 0.5~K (left) and 2.0~K (right).
Figure is from Ref.~[\onlinecite{TJH_JPSJ_2012}].}
\end{center}
\end{figure*}

The higher-temperature magnetisation curves for the polycrystalline samples of these compounds have been reported by Karunadasa {\it et al.}~\cite{Karun} and revealed non-linear behaviour of magnetisation in field with pronounced maxima in the derivatives, $dM/dH$ as a function of applied field.
Further single crystal magnetisation~\cite{TJH_JPSJ_2012} and heat capacity~\cite{THC_JPCM_2013} measurements, however, revealed highly anisotropic behaviour, which is partially masked in the polycrystalline samples.
We therefore summarise in this section the results obtained on single crystal samples.

Magnetisation versus field curves $M(H)$ and their field derivatives $dM/dH$ obtained for \sho\ for a field applied along the principal symmetry axes are shown in Fig.~\ref{Fig6_SHO_MvsH}.
For $H \! \parallel \! a$ (which is a hard magnetisation direction), $M(H)$ remains rather small in any field.
For other two directions of an applied field, a significant portion of the total magnetic moment is recovered, although no complete saturation of magnetisation is observed, as the $dM/dH$ values remain nonzero even in a field of ~70~kOe~\cite{TJH_JPSJ_2012}.
This implies that the spins of the Ho$^{3+}$ ions are not fully aligned at this field.
For $H \! \parallel \! b$ the magnetisation process is characterised by a double phase transition (seeing most clearly as two maxima in the $dM/dH$ curves in the bottom-left panel of Fig.~\ref{Fig6_SHO_MvsH}) indicative of the appearance of magnetisation plateau.
Although the plateau is not well-pronounced, i.e. the derivative $dM/dH$ remains positive and relatively large between the maxima, one has to remember that the temperature for these measurements was relatively high, 0.5~K.
For $T=2.0$~K (see right-hand panels in Fig.~\ref{Fig6_SHO_MvsH}), the plateau in the magnetisation disappears.
Therefore, it is likely that at the temperatures approaching 0~K the plateau will be much more obvious.
The value of the magnetisation on the plateau is about a third of the value observed in higher field.
This fact allowes to conjecture~\cite{TJH_JPSJ_2012} that the magnetic structure on the plateau is of collinear {\it up-up-down} type, where on each triangle the two moments are pointing along the applied field and the third moment is antiparallel to them. 

For $H \! \parallel \! c$ the magnetisation process in \sho\ is characterised by a single, relatively sharp phase transition, above which the magnetisation remains practically constant.
For $T=2.0$~K (see right-hand panels in Fig.~\ref{Fig6_SHO_MvsH}), the plateau in the magnetisation disappears, the maximum in $dM/dH$ for $H \! \parallel \! c$ broadens and shifts to slightly higher fields, while the $M(H)$ curve for $H \! \parallel \! a$ remains unchanged.

In \seo\ and \sdo\ the field dependence of the magnetisation looks rather similar to what have been observed in \sho, but the directions of applied field along which the plateaux and sharp single phase transitions appear (as well as the actual values of critical fields) are different.
In all three compounds a single and relatively sharp increase in magnetisation is seen for $H \parallel c$, a direction along which the magnetic moments are pointing in the ${\bf k}=0$ structures of \seo\ and \sho.
In \seo\ the application of field along the $a$ axis results in a magnetisation plateau, while the $b$ axis seems to be a hard magnetisation direction.
In \sdo\ a magnetisation plateau appears for $H \parallel b$, while the $a$ axis seems to be a hard magnetisation direction. 
\begin{figure}[tb]
\begin{center}
\includegraphics[width=0.8\columnwidth]{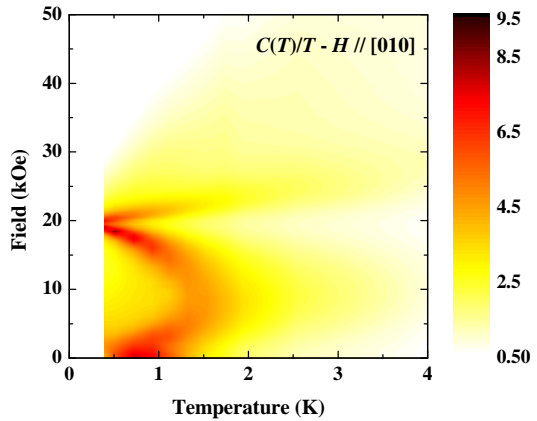}
\caption{\label{Fig7_SDO_PhD}
Magnetic $H-T$ phase diagram of \sdo\ for $H \parallel [010]$ obtained from the heat capacity measurements.
Colour represents the heat capacity divided by temperature in the units of J/(mol K$^2$).
Figure is from Ref.~[\onlinecite{THC_JPCM_2013}].}
\end{center}
\end{figure}

The fact that in all three compounds the observed plateaux in magnetisation appear at approximately a third of the magnetisation saturation values suggests that for the field applied along either the $a$ or $b$ axes the contribution from the ${\bf k}=0$ structures remains rather weak, that is, the magnetic moments in these structures remain pointing along the $c$ axis.

Further insight into the field-induced properties of \sdo\ can be gained from the heat capacity measurements~\cite{THC_JPCM_2013}.
Fig.~\ref{Fig7_SDO_PhD} (where the value of the heat capacity divided by temperature is represented by the colour scale shown on the right of the figure) shows a magnetic phase diagram of \sdo\ obtained by combining the heat capacity field-scans for $H \parallel b$.
At the lowest experimentally available temperature of 0.39~K, a sharp double peak at about 20~kOe is the main feature in the $C(H)$ curve.
The peaks indicate multiple magnetic field-induced transitions in \sdo\ for this direction of an applied field, but from the bulk-property measurements alone it is impossible to determine whether any of the field-induced phases are long-range in nature.
Therefore further neutron diffraction experiments are required to answer this question.
Remarkably for $H \parallel c$ the application of a magnetic field does not result in any features in the $C(H)$ curves sharp enough to be indicative of a phase transition~\cite{THC_JPCM_2013} which emphasises once again the highly anisotropic nature of the magnetisation process in the \slo\ compounds.

\section{Further considerations}
\subsection{Other \slo\ compounds}
\begin{figure}[tb]
\begin{center}
\includegraphics[width=0.8\columnwidth]{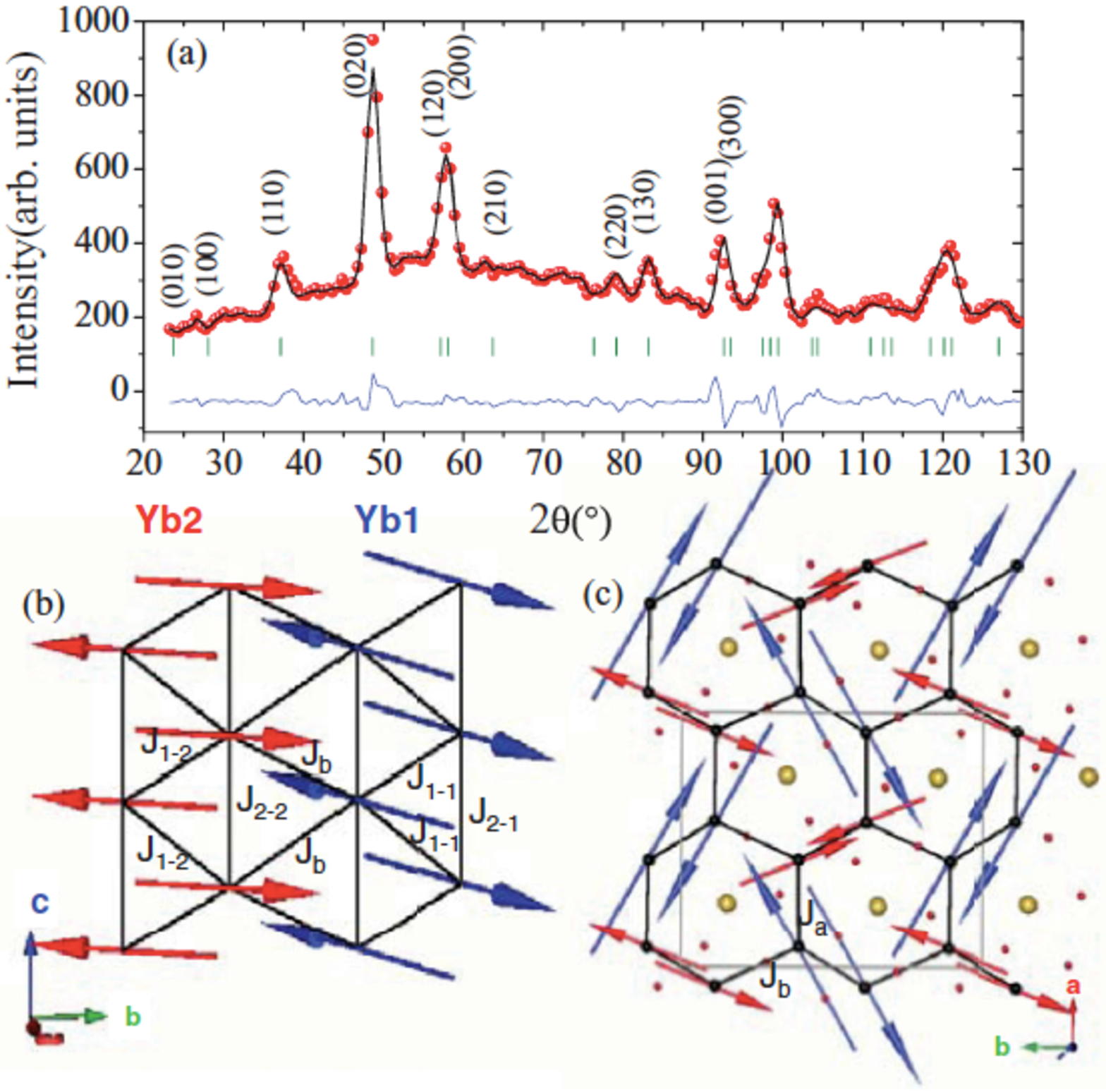}
\caption{\label{Fig8_SYO_diff}
(a) Magnetic powder pattern of \syo\ collected on D7 at 30~mK.
The magnetic structure where the arrows represent the Yb$^{3+}$ ions spins (Yb1 blue, Yb2 red) (b) along the zigzag chains and (c) projected onto the ab plane.
The Sr$^{2+}$ and O$^{2-}$ ions are represented by yellow and red circles, respectively.
Figure is from Ref.~[\onlinecite{Castro}].}
\end{center}
\end{figure}

\begin{figure}[tb]
\begin{center}
\includegraphics[width=0.8\columnwidth]{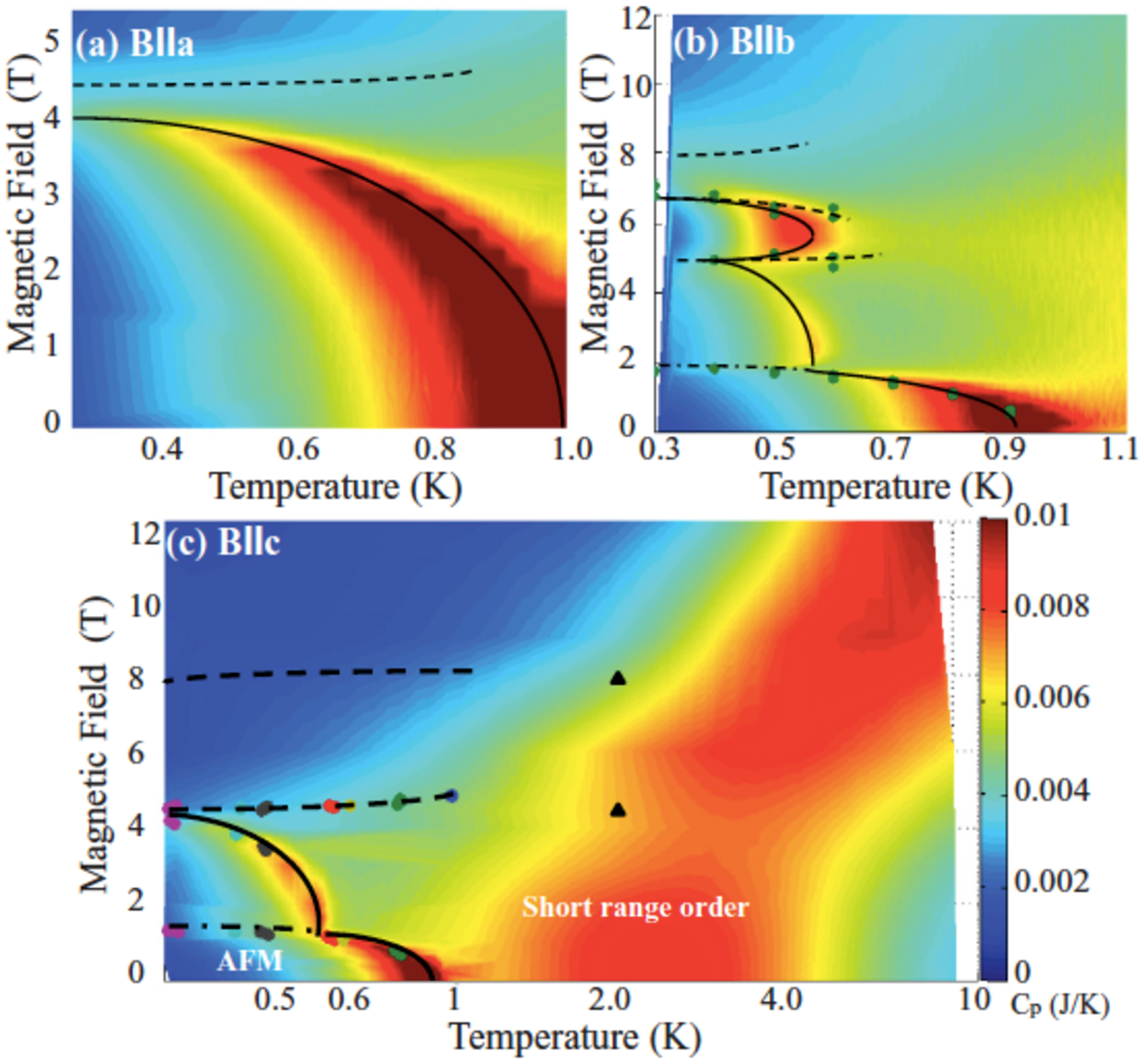}
\caption{\label{Fig9_SYO_PhD}
Magnetic phase diagram of \syo\ with magnetic field along (a) $a$, (b) $b$, and (c) $c$ axes.
The colours indicate the heat capacity in units of $J$/K.
The circles indicate the critical fields extracted from magnetocaloric effect measurements and the the triangles the critical fields extracted from magnetisation measurements.
Black solid lines show second-order phase transitions.
Dash-dot black lines indicate the transition from the AFM phase to a less ordered one.
Dashed black lines show metamagnetic crossovers.
For the phase diagram along the $c$ axis, above 4.5~T there are just three heat capacity scans at 6, 9, and 12~T, the colours between them result from the interpolation of the data.
Below 4.5~T, the data were collected every 0.2~T.
The temperature axis is in logarithmic scale.
Figure is from Ref.~[\onlinecite{Castro}.]}
\end{center}
\end{figure}

Apart from \seo, \sho\ and \sdo\ reviewed above, the only other family-member for which the low-temperature properties have been reported is \syo.
Heat capacity measurements~\cite{Castro} revealed a magnetic phase transition to LRO at $T_N=0.92$~K.
Neutron diffraction measurements~\cite{Castro} (see Fig.~\ref{Fig8_SYO_diff}a) showed that the structure is a noncollinear ${\bf k}=0$ antiferromagnet in which the magnetic moments of two inequivalent Yb$^{3+}$ ions lie in the $ab$ plane, but have different moment sizes and directions.
Both moments are reduced from the fully ordered moment of Yb$^{3+}$ (see Figs. ~\ref{Fig8_SYO_diff}b and~\ref{Fig8_SYO_diff}c).
Similarly to what has been observed for other \slo\ compounds, the application of a relatively strong field, 140~kOe, along any direction does not result in a recovery of a full moment expected for the Yb$^{3+}$ ions~\cite{Castro}. 

Very interesting and highly anisotropic magnetic phase diagrams (see Fig.~\ref{Fig9_SYO_PhD}) have been reconstructed for \syo\ from the magnetocaloric and the magnetisation measurements~\cite{Castro}.
A large number of transitions and crossovers were found which has been taken as an indication of the presence of metamagnetic phases due to spin-flip and spin-flop processes as well as possible competition between exchange interactions and magnetic anisotropy, however, the exact nature of the field-induced phases in \syo\ remains presently unknown.

We have recently started investigations of the magnetic properties of \sgo\ and from the heat capacity measurements have found~\cite{Young_unpublished} that the compound undergoes two magnetic transitions at 2.72 and 0.47~K.
The initial characterisation of \sgo\ by Karunadasa {\it et al.}~\cite{Karun} missed the higher temperature transition.
Since Gd$^{3+}$ is expected to be nearly isotropic, it is quite surprising that the transition temperature of \sgo\ is much higher than that of the other members of the family.
The higher ordering temperature of \sgo\ is, however, consistent with the properties of structurally similar $\rm Ba{\it Ln}_2O_4$ family, in which $\rm BaGd_2O_4$ orders at 2.6~K, while the rest of the compounds do not order down to at least 1.7~K\cite{Doi}.
In the absence of neutron diffraction data the magnetic structure of \sgo\ remains unknown at present.
The only other established fact about low-temperature properties of \sgo\ is that it undergoes a field-induced transformation at 20.5~kOe (for $T=0.48$~K) in a field applied along the $c$ axis~\cite{Young_unpublished}.

It would certainly be interesting to expand the \slo\ family and to test the magnetic properties of Tm, Tb, Sm, Nd containing compounds provided that single crystal samples can be prepared.

\subsection{Crystalline electric field effects}
From the findings presented above for the \slo\ compounds, which vary greatly from one $Ln$ ion to another, it is rather obvious that the low-lying crystalline electric field (CEF) levels must play an important role in the formation of the highly anisotropic magnetic properties.
At present the CEF schemes remain unknown and the task of establishing them may not be trivial: there are 8 $Ln$ ions on two district crystallographic sites in the unit cell.
The symmetry is rather low, therefore the number of CEF levels is expected to be large.
Also, the positions of the levels at lower temperature can potentially be influenced by the development of short-range magnetic order.
Inelastic neutron scattering (INS) data for \sdo\ and \sho\ have been collected back in 2005 by Kenzelman {\it et al.}~\cite{Kenzelman}.
The more recent INS results reported for \sho\ by Ghosh {\it et al.}~\cite{Ghosh} are largely in agreement with the previous data.
We have also collected further INS data for \seo \cite{CEF_Hayes}, but to date neither group have reported any CEF schemes.
Moreover, there are further indications~\cite{CEF_Desilets} that the problem could prove to be difficult to solve.
An additional motivation for preparing this review was to alert the frustrated magnetism community to the presence of such a challenging, but potentially very important problem. 

\section{Summary}
\begin{table*}
\begin{center}
\begin{tabular}{| c | r | c | c | c | l |}
\hline
$Ln$	& $\Theta_{W}$, K	&$p \rm _{eff}$, $\mu_B$& $T_N$, K				& Magnetic structure (in zero field)					& $H_C$, kOe											\\	\hline
	& 				&				& 							& ${\bf k}=0$ LRO AFM (moments $\parallel$ $c$ axis)\cite{OAP_PRB_2008}	&$H \! \parallel \! c$: 5.4 (0.5~K)\cite{TJH_JPSJ_2012}		\\ 
Er	& -13.5~\cite{Karun}	&9.176~\cite{Karun}	& 0.75~\cite{OAP_PRB_2008}		& \& 											& 													\\	
	& 				& 				& 							& quasi 1D SRO AFM (moments $\parallel$ $a$ axis)\cite{TJH_PRB_2011}	& $H \! \parallel \! a$: 2.0 \& 12.5 (0.5~K)\cite{TJH_JPSJ_2012}	\\	\hline
	& 				&				&							& ${\bf k}=0$ SRO AFM  (moments $\parallel$ $c$ axis)	&  $H \! \parallel \! c$: 4.0 (0.5~K)\cite{TJH_JPSJ_2012}	\\ 
Ho	& -16.9~\cite{Karun}	& 10.50~\cite{Karun}	& 0.68~\cite{OY_JPCS_2012}		&  \& 											& 										 	\\ 	
	& 				& 				& 							&  quasi 1D SRO AFM (moments $\parallel$ $b$ axis) \cite{OY_PRB_2013}	& $H \! \parallel \! b$: 5.9 \& 12.0 (0.5~K)\cite{TJH_JPSJ_2012}	\\	\hline
Dy	& -22.9~\cite{Karun}	& 10.35~\cite{Karun}	& $<$0.02~\cite{THC_JPCM_2013} 	& only SRO above 0.02~K~\cite{THC_JPCM_2013}		& $H \! \parallel \! b$: 1.6 \& 20.3 (0.5~K)\cite{TJH_JPSJ_2012}, 20 (0.39~K)\cite{THC_JPCM_2013} \\ 	
	& 				& 				& 							& 											& $H \! \parallel \! c$: 12.0 (0.5~K)\cite{TJH_JPSJ_2012}			\\	\hline
Gd	& -9.0~\cite{Karun}	& 8.02~\cite{Karun}	& 0.47 \& 2.72~\cite{Young_unpublished}	 	& unknown							& $H \! \parallel \! c$: 20.5 (0.48~K)\cite{Young_unpublished}		\\	\hline
	& 				& 				& 							& noncollinear ${\bf k}=0$ AFM						& $H \! \parallel \! a$: 45 (1.0~K)\cite{Castro}					 \\
Yb	& -99.4~\cite{Karun}	& 4.348~\cite{Karun}	& 0.92~\cite{Castro}				& with different moment sizes and directions\cite{Castro}	& $H \! \parallel \! b$: 15 \& 60 (0.6~K)\cite{Castro}				 \\ 	
	& 				& 				& 							& 											& $H \! \parallel \! c$: 11 \& 45 (0.6~K)\cite{Castro}				\\ \hline
\end{tabular}
\caption {\label{Table1} Summary of the magnetic properties of the \slo\ compounds.
$\Theta_{W}$ is Weiss temperature and $p \rm _{eff}$ is an effective magnetic moment.
Both parameters were determined from the higher-temperature susceptibility measurements.
For the transitions fields $H_C$ the corresponding measurement temperatures are indicated in the brackets. 
}
\end{center}
\end{table*}
We conclude this review by listing in Table~\ref{Table1} the most important magnetic parameters of the \slo\ compounds, such as Weiss temperature $\Theta_{W}$, effective moment $p \rm _{eff}$ in Bohr magnetons $\mu_B$, magnetic ordering temperature $T_N$ as well as indicating the nature of the zero field ground state and the presence of critical fields $H_C$ for different directions of an applied field.

Important pieces of information missing from Table~\ref{Table1} include the values of the various exchange interactions and details on the magnetic anisotropy in the \slo\ compounds.
This information which is typically obtained from inelastic neutron scattering experiments is so far unavailable.
Only after establishing the absolute values (including signs) and relative strengths of the relevant exchange interactions, as well as details of the magnetic anisotropy, could one classify the \slo\ compounds as a collection of weakly interacting chains of magnetic moments, or as a network of ladders consisting of triangles.
Apart from neutron scattering, further Monte Carlo simulations, both direct and reverse, as well as density-functional theory band-structure calculations may play an important role in determining the magnetic interactions.

In \seo\ the LRO ${\bf k}=0$ AFM phase (see Fig.~\ref{Fig3_SEO_str}) in which the magnetic moments point along the $c$ axis appears below 0.75~K while in \sho\ a very similar phase appearing below 0.68~K remains short-range ordered down at least 50~mK.
Apart from this phase, a SRO quasi one-dimensional AFM component is found in both compounds, but the direction along which the spins are pointing is different - it is parallel to the $a$ axis in \seo\ and parallel to the $b$ axis in \sho.
In \seo\ the diffuse scattering signal corresponding to the quasi 1D component appears to be much more structured compared to the \sho, which could be indicative of the importance of further neighbour exchange interactions.
In \syo\ the magnetic moments are confined to the $ab$ plane (see Figs.~\ref{Fig8_SYO_diff}b and \ref{Fig8_SYO_diff}c), with the two different Yb$^{3+}$ sites having very different moment sizes and directions~\cite{Castro}.
No long-range magnetic order has been found in \sdo\ down to 20~mK.
Despite having the weakest magnetic interactions (as demonstrated by the lowest Weiss temperature) \sgo\ orders at the highest temperature of 2.72~K and undergoes another transition at 0.47~K.
This observation suggests an immense importance of the magnetic anisotropy in establishing the ground state of the \slo\ compounds and the potential competition between the exchange interactions and the single-ion effects.

For all the \slo\ compounds the application of an external magnetic field results in the appearance of complex and highly anisotropic phase diagrams revealing multiple phase transitions, magnetisation plateaux and cross-over regions.
The magnetic structure of the field-induced phases remains presently unknown.

We hope that this review will stimulate further research on the magnetic properties of the \slo\ and related honeycomb lattice compounds.

\section{Acknowledgements}
The author is very grateful to G.~Balakrishnan, M.R.~Lees, D.McK.~Paul, N.R.~Wilson, T.J.~Hayes, O.~Young, T.H.~Cheffings, 
P.~Manuel, D.D.~Khalyavin, D.T.~Adroja, F.~Demmel, B.D.~Rainford, 
A.R.~Wildes, B.~Ouladdiaf, L.C.~Chapon, S.~de~Brion, E.~Suard, C.~Ritter, P.P.~Deen, 
R.J.~Mason, A.K.R.~Briffa, M.W.~Long, 
J.~Mercer, M.L.~Plumer, A.~Desilets-Benoit, A.D.~Bianchi, 
and D.L.~Quintero-Castro for the collaborations on the projects involving the magnetic properties of the \slo\ compounds.

\end{document}